\documentclass[aps,twocolumn,floats,superscriptaddress]{revtex4-1}

\usepackage{graphicx}
\usepackage{amssymb,amsmath}
\usepackage{color,rotating,soul}

\newcommand{\Pe}{\ensuremath{\text{Pe}}}

\newcommand{\beginsupplement}{%
\setcounter{table}{0}
\renewcommand{\thetable}{S\arabic{table}}%
\setcounter{figure}{0}
\renewcommand{\thefigure}{S\arabic{figure}}%
\setcounter{equation}{0}
\renewcommand{\theequation}{S\arabic{equation}}%
}

\begin{document}

\title{A self-driven phase transition drives \emph{Myxococcus xanthus} fruiting body formation}

\author{Guannan Liu}
\email{These authors contributed equally to this work.}
\affiliation{Joseph Henry Laboratories of Physics and Lewis-Sigler Institute for Integrative Genomics, Princeton University, Princeton, NJ 08544, USA}

\author{Adam Patch} 
\email{These authors contributed equally to this work.}
\affiliation{Department of Physics and Soft and Living Matter Program, Syracuse University, Syracuse, NY 13244, USA}

\author{Fatmag\"ul Bahar}
\email{These authors contributed equally to this work.}
\affiliation{Department of Biology, Syracuse University, Syracuse, NY 13244, USA}

\author{David Yllanes}
\email{These authors contributed equally to this work.}
\affiliation{Department of Physics and Soft and Living Matter Program, Syracuse University, Syracuse, NY 13244, USA}
\affiliation{Instituto de Biocomputaci\'on y F\'isica de Sistemas Complejos (BIFI), 50009 Zaragoza, Spain}

\author{Roy D. Welch}
\affiliation{Department of Biology, Syracuse University, Syracuse, NY 13244, USA}

\author{M. Cristina Marchetti}
\affiliation{Department of Physics and Soft and Living Matter Program, Syracuse University, Syracuse, NY 13244, USA}

\author{Shashi Thutupalli}
\email{Present address: Simons Center for the Study of Living Machines, National Centre for Biological Sciences, Tata Institute for Fundamental Research, Bangalore 560065, India}
\affiliation{Joseph Henry Laboratories of Physics and Lewis-Sigler Institute for Integrative Genomics, Princeton University, Princeton, NJ 08544, USA}

\author{Joshua W. Shaevitz}
\email{E-mail: shaevitz@princeton.edu}
\affiliation{Joseph Henry Laboratories of Physics and Lewis-Sigler Institute for Integrative Genomics, Princeton University, Princeton, NJ 08544, USA}

\date{\today}

\begin{abstract}
Combining high-resolution single cell tracking experiments with numerical simulations, we show that starvation-induced fruiting body (FB) formation in \emph{Myxococcus xanthus} is a phase separation driven by cells that tune their motility over time. The phase separation can be understood in terms of cell density and a dimensionless P\'eclet number that captures cell motility through speed and reversal frequency. Our work suggests that \emph{M. xanthus} take advantage of a self-driven non-equilibrium phase transition that can be controlled at the single cell level.
\end{abstract}

\maketitle

Unicellular organisms such as bacteria and amoeba are capable of spontaneously organizing into complex multicellular structures~\cite{laub1998molecular,dormann2002becoming}. A striking example of such collective behavior is the starvation-induced organization of the rod-shaped, soil-dwelling bacterium \textit{Myxococcus xanthus} into macroscopic, multicellular aggregates known as ``fruiting bodies''~(FBs)~\cite{zusman2007chemosensory}. When nutrients are scarce, {\it M. xanthus} cells undergo a multicellular process of self-organization during which cells move to form dome-shaped aggregates comprising hundreds of thousands of cells. A subset of cells at the center of each droplet differentiate to form metabolically quiescent spores that can survive long periods of starvation~\cite{zusman2007chemosensory,starruss2012pattern,shimkets1990social}. 

The striking phenotypic similarity between FB formation in {\it M. xanthus} and in the amoeba {\it Dictyostelium discoidium} has led to the longstanding hypothesis that {\it M. xanthus} FB formation is driven by long-range chemical signaling mechanisms, as it is in the amoeba. Although {\it M. xanthus} cells are thought to employ chemical communication to initiate FB formation (termed A-signaling)~\cite{kuspa1992identification}, to synchronize reversal frequency (termed C-signaling)~\cite{lobedanz2003identification,shimkets1990csga}, and to communicate through mucopolysaccharide ``slime trails'' that other cells can sense and follow~\cite{burchard1982trail}, a quantitative understanding of the mechanisms that drive aggregation has remained elusive. 

\emph{M. xanthus} cells move by gliding on solid surfaces using both tank-tread-like transport motors and the retraction of extruded filaments called pili, and can modulate their speed in a continuous manner~\cite{balagam2014myxococcus,hodgkin1979genetics}. The cells also have the ability to reverse their direction of motion, typically every several minutes, and can modify the reversal frequency in different situations~\cite{wu2009periodic,blackhart1985frizzy,thutupalli2015directional}. In addition to the role of chemical signaling, previous modeling work has attempted to investigate mechanical aspects of FB formation by considering contact-mediated interactions between cells, although these ideas have not been thoroughly tested experimentally~\cite{sozinova2005three}.

Using experiments and insight from theory, we demonstrate that \emph{M. xanthus} FB formation can be described as a phase separation process driven, at least initially, by changes to the motility of individual cells. Importantly, this appears to happen in the absence of complex signaling mechanisms and interactions between cells, and requires no real-time control at the cellular level. While the ability to actively change motility ultimately leads to a phase transition, cells do not have to implement a complicated feedback mechanism to alter motility in response to specific chemical or mechanical cues. Rather, cells need only speed up and suppress reversals upon starvation and the collective mechanics then naturally induces phase separation of the entire population. The theoretical inspiration for this work is the Motility-Induced Phase Separation (MIPS), where purely repulsive Active Brownian Particles (ABPs) spontaneously aggregate through a jamming-based phase transition~\cite{Fily2012a,Cates2015,Marchetti2016a}. While there are important differences between the MIPS  and the droplet formation seen in {\em M. xanthus} populations, we use this model to explore parameters and regions of phase space that are inaccessible to experiments.

\begin{figure}[t]
\centering
\includegraphics[width=1\columnwidth]{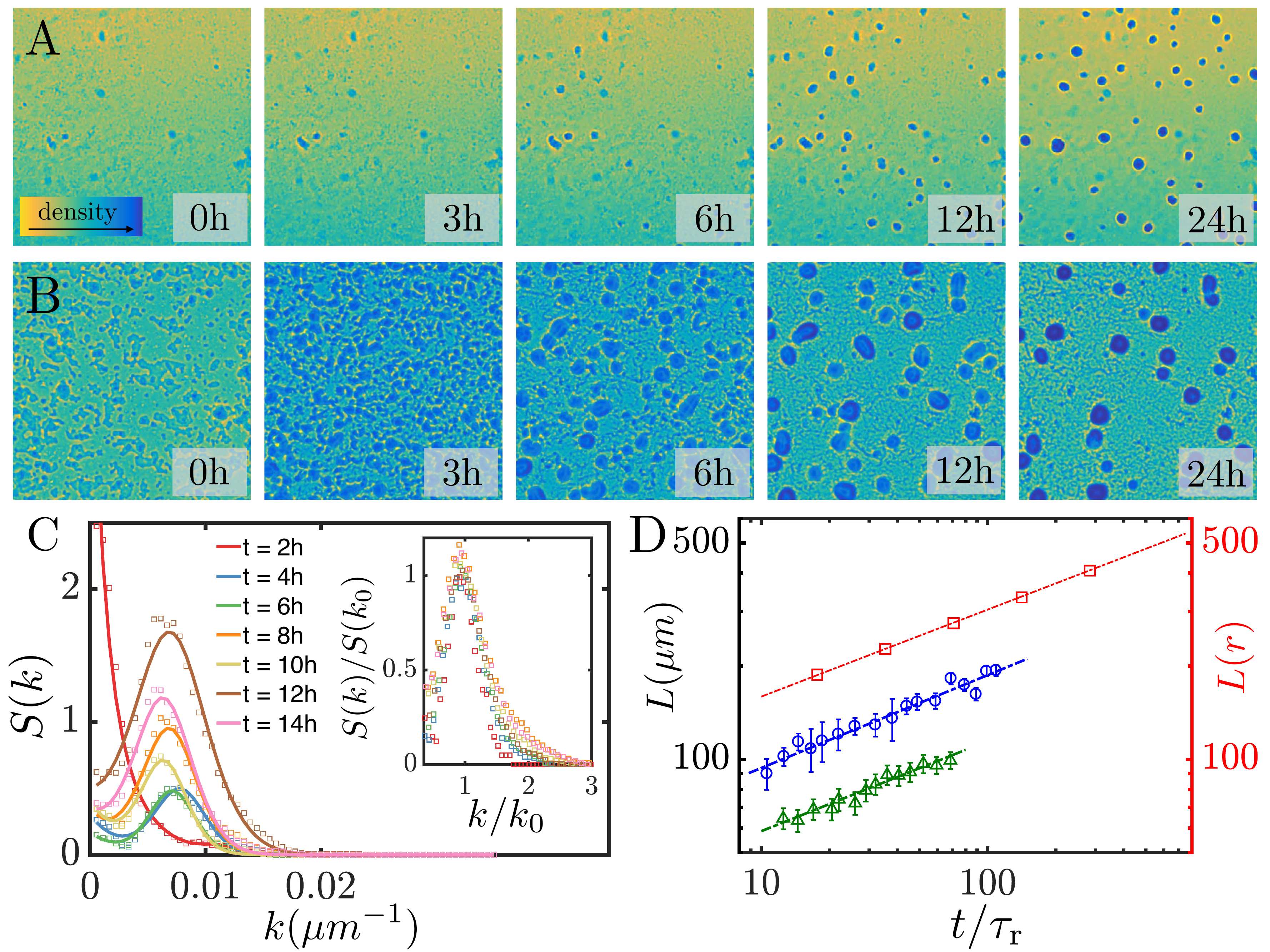}
\caption{Aggregation in \emph{M. xanthus}. (A, B) Microscopy images of {\it M. xanthus} cells undergoing phase separation via a nucleation and growth process  (A, low density at 1.5$\times$10$^{8}$ cell/mL) and spinodal decomposition (B, high density at 5$\times$10$^{8}$ cell/mL). Each image has dimension 1.5~mm by 1.5~mm. (C) The structure factor at different spatial frequencies for {\it M. xanthus} phase separation at high density and different times after starvation is calculated as the magnitude squared of the radial component of the Fourier transform of images. Solid lines are fits to the sum of a Gaussian function and a decaying exponential. Time $t=0$ denotes the first frame in a movie where macroscopic coarsening is observed. (C, inset) Collapse of the radial Fourier transform when frequency is scaled by the peak wavenumber $k_0$ and the amplitude is scaled by the peak amplitude $S(k_0)$. (D) Power law scaling of the dominant length scale with time, $L\sim t^\alpha$, for nucleation and growth (green triangles, $\alpha=0.29 \pm 0.02$) and spinodal decomposition (blue circles, $\alpha=0.30 \pm 0.02$) experiments (left axis), and ABP spinodal decomposition simulations (red squares, $\alpha=0.281 \pm 0.002$, right axis). Time is written in units of the reversal time, $\tau_\text{r}\approx10$~min. Error bars represent one standard-error.
\label{fig:SnapShots}}
\end{figure} 

We first investigated the dynamics of FB formation using different cell densities (experimental details are described in the Supplemental Materials, SM). Time-lapse, bright-field images were used to quantify the resultant dynamics (Fig. \ref{fig:SnapShots}, Movies S1-3). Pixel intensity is indicative of local population density with darker regions corresponding to the FBs and the lighter pixels corresponding to the low density regions of bacteria. When the inoculation cell density is very low (2.5$\times$10$^{7}$~cells/mL), no large-scale structure formation is seen (Movie S1). Over the first few hours, cells largely move independently, reversing frequently and with minimal cell-cell contacts and interactions. This results in the formation of spatially stable nematic streams at later times ($\sim$ 6-8 hours) but not fruiting bodies ~\cite{thutupalli2015directional}.

When the inoculation density is increased to 1.5$\times$10$^{8}$~cells/ml, FBs form randomly in space and time through the experiment (Fig.~\ref{fig:SnapShots}A, Movie S2). In a field of view of 3~mm by 2.5~mm, approximately 10 FB droplets were observed after 24 hours, although in some cases as few as 2 droplets formed. This spatio-temporally random appearance of FBs is similar to a phase separation process called nucleation and growth in which an energy barrier between two phases causes small fluctuations in the population density to die out. The later stages of coarsening involve significant flux between neighboring FBs, seen directly in some experiments where the cell movement is evident and in others where small droplets are observed to dissolve into nearby larger ones. This is reminiscent of an Ostwald ripening process and recent work used a model of Ostwald ripening to predict the disappearance and persistence of {\it M. xanthus} FBs~\cite{bahar2014describing}.

When the inoculation density was further increased to 5$\times$10$^{8}$ cells/mL and above, we observed that FBs formed via a different dynamical mechanism (Fig.~\ref{fig:SnapShots}B, Movie S3). Rather than spatially random nucleation and slow growth, high-density cultures spontaneously and immediately begin to phase separate over the entire field of view. Within the first 11 hours after plating (Fig. \ref{fig:tauHist}), we observed the formation of a global instability in the cell density that resulted in small, mesh-like structures that cover the petri dish. This kind of spontaneous phase separation, similar to spinodal decomposition, classically arises when microscopic fluctuations in the local density are inherently unstable, lacking an energy barrier to separate the homogeneous and more favorable phase-separated regimes. As the mesh coarsened over time, small droplets appeared that were connected by thinner layers of cells. Finally, a subset of these droplets grew and turned into round FBs. We determine if an experiment exhibits phase separation and the underlying kinetic mechanism by analyzing the temporal dynamics in the image as described in the SM.

We next compared the dynamics of FB formation to the well-studied MIPS seen in simulations of ABPs. Briefly, we simulated particles moving with speed $v_0$ along a direction that is randomized at rate $D_r$ and reversed at rate $f_\text{rev}$ (see SM for further details). As has been shown previously, and similar to our observations from {\it M. xanthus} cells, ABPs aggregate in a density-dependent manner~\cite{Fily2012a,Cates2015,Marchetti2016a}. At very low densities, no aggregation is observed (Movie S4), whereas the simulations produce nucleation and growth and finally spinodal decomposition as the density of particles is increased (Movies S5 and S6). In this model, aggregation occurs as particle-particle collisions cause  jamming that can only be relieved though rotation of the orientation vectors. 

A hallmark of spinodal decomposition is a well-defined length scale of the phase-separated domains that increases with time as a power-law as the domains grow self-similarly~\cite{bray2002theory,chaikin2000principles}. At high inoculation densities corresponding to the spinodal regime, a single dominant length scale emerges in the organization of the bacterial domains (Fig. \ref {fig:SnapShots}C) and grows in time as a power law with an exponent $\alpha = 0.30 \pm 0.02$ (Fig. \ref{fig:SnapShots}D). When the structure factor $S(k)$ is rescaled by the  peak wavenumber and amplitude, the shape of the peak remains largely constant over time up to 14 hours, representing self-similar coarsening of the phase separated domains (Fig. \ref {fig:SnapShots}C inset)~\cite{marro1979,bray2002theory}
\footnote{ $S(k)$ is typically normalized by the square of the peak wavenumber for a 2D system, but nonlinearities in our imaging and lighting make this unfeasible.}
. At later times, the initially mesh-like domains break up into rounder droplets in what appears to be a separate part of the phase separation mechanism. A similar increase in the dominant length scale with time is seen at low inoculation densities where the size of the individual aggregates in the nucleation and growth regime grows in time with the same exponent, $\alpha = 0.29 \pm 0.02$. We next measured the length scale for ABPs undergoing a MIPS and recover a similar scaling, $\alpha = 0.281\pm0.002$
\footnote{Both motility-induced and equilibrium phase separation show a crossover from a faster growth law at short times, corresponding to the initial nucleation of clusters, to the slower coarsening regime shown in Fig. 1D. This short-time regime is not, however, accessible to experiments. The full kinetics of MIPS has been studied in~\cite {Patch2017}.}
. These results are in agreement with previous results for non-reversing ABPs~\cite{Stenhammar2013a,Redner2013b,Patch2017} and continuum models of MIPS~\cite{Wittkowski2014}. Interestingly, we find that the time evolution of the characteristic size is the same for active 2D aggregation (MIPS simulation) and {\it M. xanthus} FB formation.

We next sought to investigate the effect of changes in motility on FB formation. The total activity in self-propelled systems can be quantified using the dimensionless inverse rotational P\'eclet number, given by the ratio of the cell size $\ell_\text{c}$ to the persistence length of the motility paths $\ell_\text{p}=v_0/D_r$~~\cite{Patch2017,Redner2013b,Stenhammar2013a,Speck2015},
\begin{equation}
\text{Pe}^{-1}_\text{r} = \ell_\text{c}/ \ell_\text{p} = \ell_\text{c} D_\text{r}/ v_0 \; .
\label{eq:Peclet}
\end{equation}
As {\em M. xanthus} cells can spontaneously reverse their direction of motion at rate $f_\text{rev}$, we sought to understand how these reversal events might change the P\'eclet number. Introducing a reversal frequency adds a new timescale, the effects of which can be incorporated in an effective rotational diffusion coefficient (see SM and \cite{grossmann2018})
\begin{equation}
D_\text{r}^\text{eff} = D_\text{r} + 2 f_\text{rev} \;.
\label{eq:Deff}
\end{equation}

\begin{figure}[t]
\includegraphics[width=1\columnwidth]{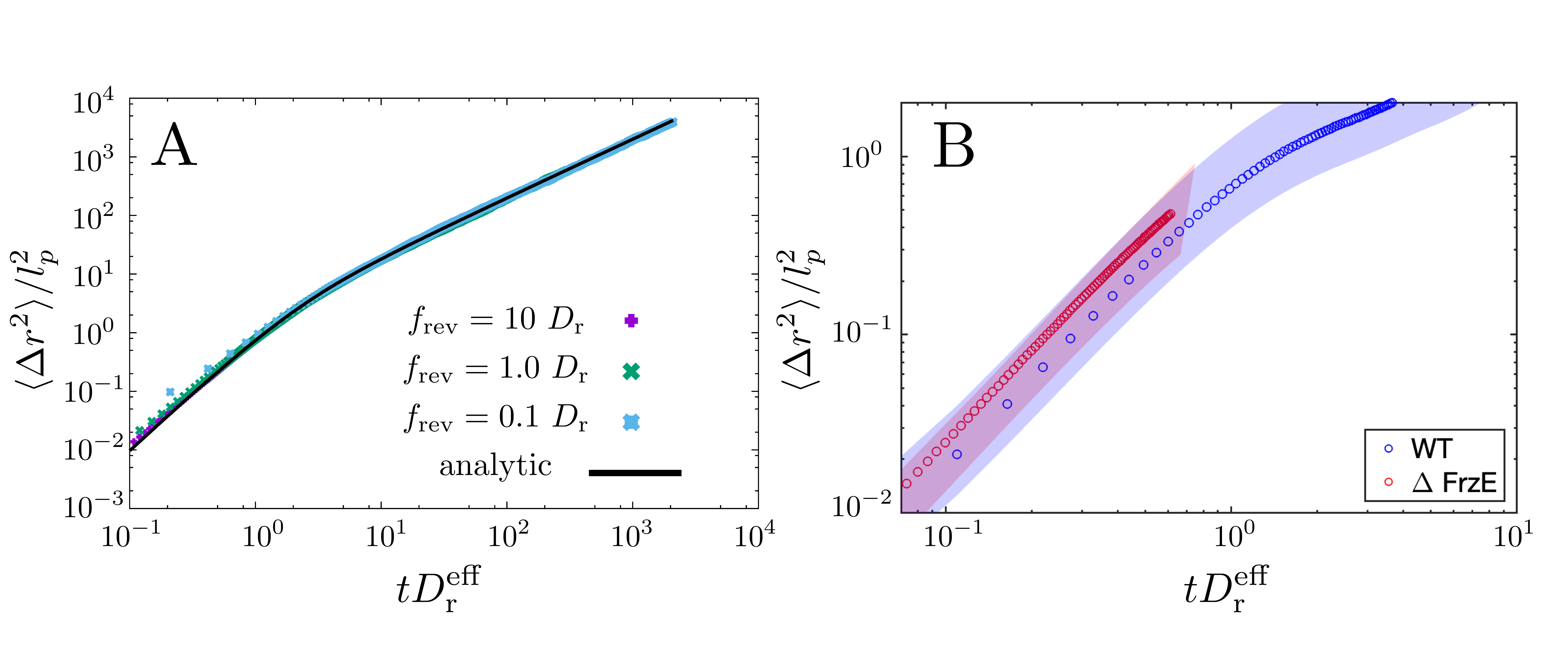}
\caption{Reversals affect rotational diffusion. (A) Mean square displacement (MSD) versus time for single particles, plotted for various reversal frequencies. The data for different reversal frequencies collapse when time is scaled by the reorientation time $\tau_\text{r}^{-1} = D_\text{r}^\text{eff}$. The     solid line is described by Eq. \ref{eq:revMSD}. (B) Experimental MSD for isolated wild-type (blue) and $\Delta$FrzE (red) {\it M. xanthus} cells. Time is scaled by $D_\text{r}^\text{eff,WT}=0.336$\ min$^{-1}$ and $D_\text{r}^\text{eff,$\Delta$FrzE}=0.065$\ min$^{-1}$, respectively. Shaded regions represent the combined standard error taking into account uncertainty in the measured speed, reversal frequency, and $D_r$.
\label{fig:Model_MSD}}
\end{figure}

We confirmed that Equation \ref{eq:Deff} accurately describes the trajectories from both simulations and moving cells. Adding reversals as a Poisson process with mean frequency $f_\text{rev}$ to a model of ABPs ~\cite{Fily2012a,Marchetti2016a,Tailleur2008}, we found that the single-particle mean squared displacement (MSD) for different reversal frequencies collapses when time is scaled by the reorientation time $\tau_\text{r}^{-1} = D_\text{r}^\text{eff}$ (Fig.~\ref{fig:Model_MSD}A). Moreover, the crossover between ballistic and diffusive motion occurs at time $t=\tau_\text{r}^{-1}$ for all $f_\text{rev}$ at both low and high particle density (Fig. ~\ref{fig:MSD_density_sim}). 

To measure the effect of reversals on trajectories of moving cells, we performed experiments at very low cell density so that the motion of individual cells was not affected by cell-cell collisions and motion was purely two-dimensional, as opposed to the three-dimensional motion in the FBs (Fig.~\ref{fig:experiment_track}A). The velocity autocorrelation function for the WT cells calculated from these data decays exponentially with small scale oscillations that die out at longer times due to the directional reversals (Fig.~\ref{fig:experiment_track}B). We fit the autocorrelation functions from both the wildtype (WT) and the non-reversing mutant $\Delta$FrzE \footnote{While $\Delta$FrzE cells have been reported to reverse at a very low frequency, we did not observe any reversals using this strain in our analysis.} and found that $D_\text{r}^\text{eff, WT} = 0.336$ min$^{-1}$ and $D_\text{r}^\text{eff, FrzE} = 0.065$ min$^{-1}$. When time is scaled by these values of $D_\text{r}^\text{eff}$, the MSD versus time plots for WT and $\Delta$FrzE cells collapse together. Interestingly, with a measured wild-type reversal frequency of 6.3~h$^{-1}$, this implies that the underlying rotational diffusion coefficient of WT cells, $D_\text{r,WT} = 0.127$ min$^{-1}$ is larger than that of $\Delta$FrzE cells, $D_\text{r,FrzE} = 0.065$ min$^{-1}$, potentially indicating the the Frz pathway may affect directionality in addition to  reversal.

\begin{figure}[t]
\centering
\includegraphics[width=1\columnwidth]{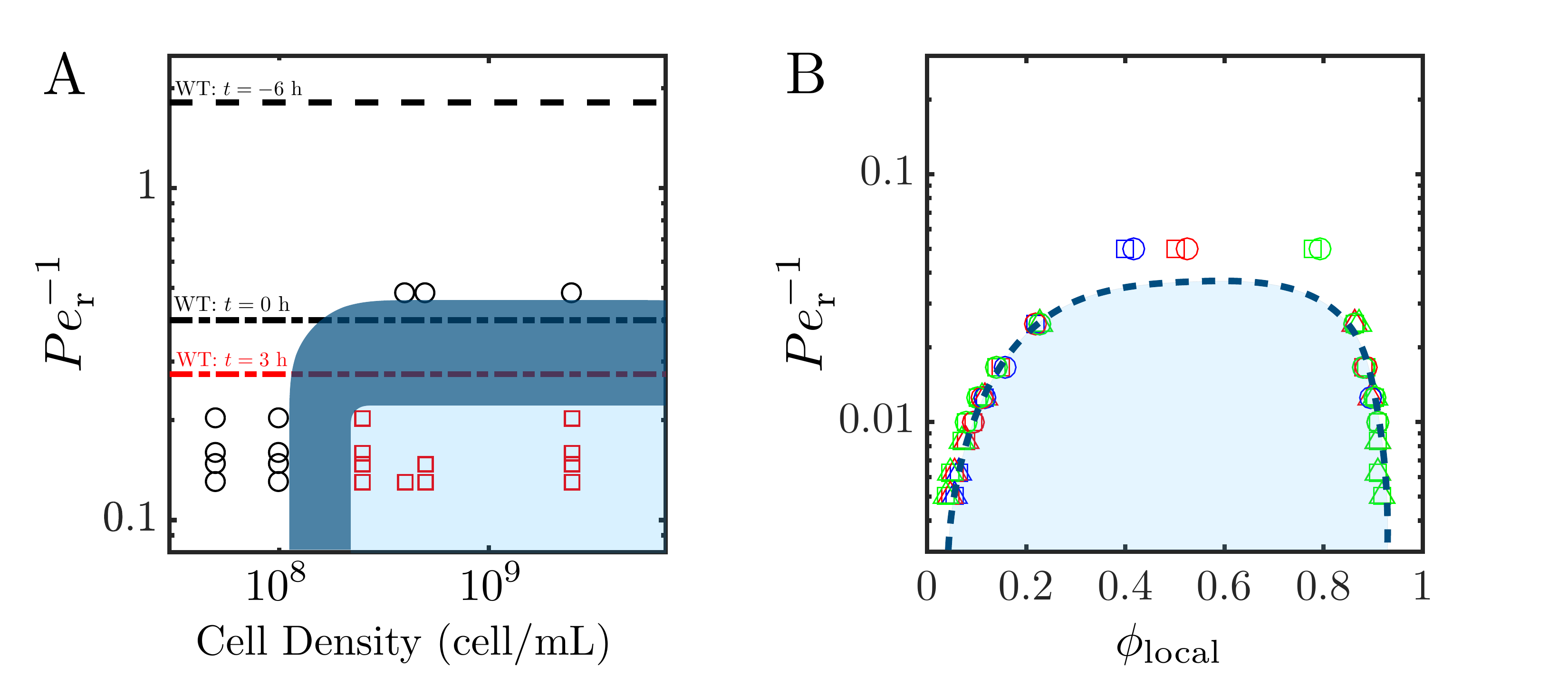}
\caption{(A) Experimental phase diagram for \emph{M. xanthus} phase separation of the non-reversing mutant $\Delta$FrzE. For each experiment at a particular density and \Pe$^{-1}_\text{r}$, we determined whether the system is phase separated via spinodal decomposition (red squares) or not phase separated (black circles) after 24 hours. The estimated region of phase space containing the spinodal line is shaded in dark blue while the spinodal region is shaded in light blue. Dashed horizontal lines denote the \Pe$^{-1}_\text{r}$ for wild-type cells one hour after inoculation ($t$=$-6$~h), at the moment coarsening is first observed ($t$=$0$~h), and 3 hours into the coarsening process ($t$=$3$~h). (B) The phase diagram for reversing ABPs showing the spinodal boundary. Spinodal points correspond to the peaks of a bimodal distribution of local density (see SM) for different values of the ratio $f_\text{rev}/D_\text{r} = 0.1\text{(triangles)}, 1\text{(squares)}, 10\text{(circles)}$ and of the packing fraction $\phi_0 = 0.45\text{(blue)}, 0.55\text{(red)}, 0.65\text{(green)}$. The dashed line is a guide to the eye. The horizontal axis is the local particle packing fraction $\phi_\text{local}$.}
\label{fig:PhaseDiagram}
\end{figure}

If FB formation is an actively driven process, changes to  \Pe$^{-1}_\text{r}$  could affect the occurrence of cellular aggregation as it does in the MIPS. To control the \Pe$^{-1}_\text{r}$ of {\it M. xanthus} experimentally, we used non-reversing $\Delta$FrzE cells and altered the propulsion speed $v_0$ using the drug nigericin~\cite{sun2011motor}. The inverse P\'eclet number for each experimental condition was estimated by separately measuring cell velocity and $D_{\text{r}}$, combined with the average cell size of $\ell_\text{c}=2.5$~$\mu$m (half a cell length). We mixed a small number of fluorescently-labeled cells with non-fluorescent cells at a ratio of 1:400 and tracked the fluorescent cells to measure their speed.  We find that the velocity of $\Delta$FrzE cells decreases monotonically from 1.25~$\mu$m/min in the absence of drug to 0.36~$\mu$m/min in the presence of 10~$\mu$M nigericin (Fig.~\ref{fig:experiment_track}C). By tracking isolated cells at very low density, we find that $D_\text{r}$ does not change when nigericin is added up to a concentration of 10~$\mu$M (Fig.~\ref{fig:experiment_track}D), in contrast to the effect in eukaryotes~\cite{maiuri2015actin}. 

We generated a phase diagram for {\it M xanthus} FB formation by performing experiments with $\Delta$FrzE cells at different inoculation densities and nigericin concentrations (Fig.~\ref{fig:PhaseDiagram}A). When starved, non-reversing cells coarsen into FBs with the same temporal scaling as WT cells, although these aggregates are unstable and short lived (Fig. \ref{fig:frzE_evolution}, Movies S7 and S8). As predicted, at low density or high \Pe$^{-1}_\text{r}$, the system does not form FBs (black circles). At high density or low \Pe$^{-1}_\text{r}$, the system phase separates via spinodal decomposition (red squares). The estimated region of the phase space that includes the spinodal line is shaded in dark blue on the phase diagram, and at high density lies between $\text{Pe}^{-1}_\text{r}=0.20$ and $0.48$. In Fig.~\ref{fig:PhaseDiagram}B, we show a similar phase diagram obtained from simulations of reversing ABPs. The scaling of the data for different values of $f_\text{rev}$ and mean density confirms that the effect of reversals can be incorporated  into $\text{Pe}^{-1}_\text{r}$ using the effective rotational diffusion coefficient $D_\text{r}^\text{eff}$. Our results are in good agreement with previous studies of ABPs without reversals~\cite{Redner2013b,Patch2017}.

\Pe$^{-1}_\text{r}$ depends on four parameters, two of which {\it M. xanthus} potentially has the ability to control during FB formation. Cells do not grow during aggregation due to the starvation conditions and $D_\text{r}$ is presumably set by thermal fluctuations of the cell body and molecular noise in the motility process. However, both the cell speed $v_0$ and the reversal frequency $f_\text{rev}$ are known to be under cellular control, suggesting that the cells might have adapted to take advantage of such control~\cite{sun2011motor,guzzo2018gated}.

\begin{figure}[t]
\centering
\includegraphics[width=.9\columnwidth]{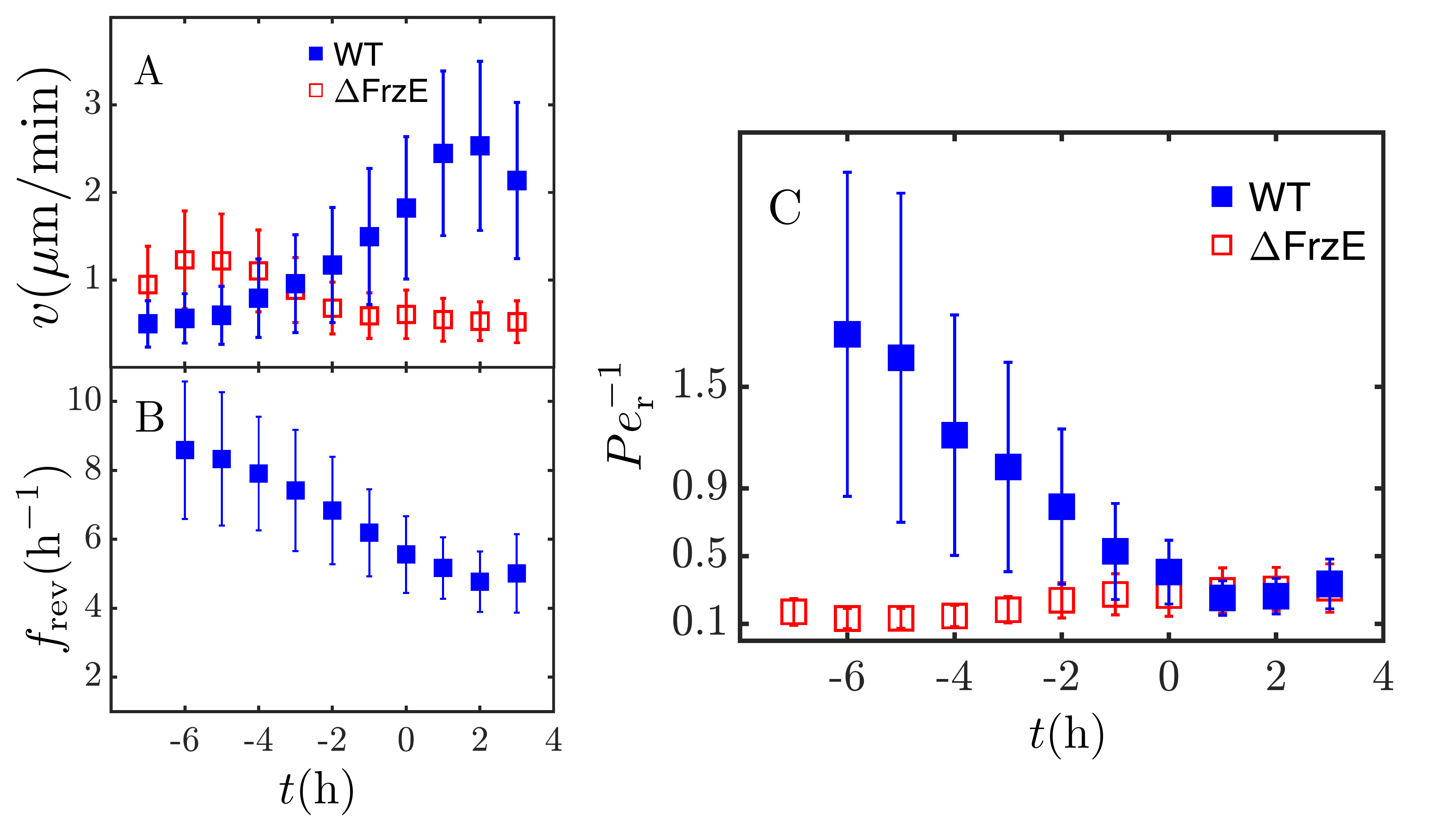}
\caption{Cell tracking over time after starvation. Cell speed (A), reversal frequency (B), and the resulting \Pe$^{-1}_\text{r}$ (C) are shown for wild-type (blue) and non-reversing $\Delta$FrzE cells (red). $\Delta$FrzE cells do not reverse, so no red points are plotted in B. Error bars represent the standard error of the mean. $t=0$ corresponds to 7 hours post inoculation, when  macroscopic coarsening was first observed in these experiments.}
\label{fig:TrackingActive}
\end{figure}

We tracked individual fluorescent wild-type cells at a population density of $5\times10^{8}$ cell/mL for 11 hours after starvation and found cells changed both their gliding speed and reversal frequency (Fig. \ref{fig:TrackingActive}). In the first 2--3 hours, cells exhibited low gliding speeds of $\sim1.3\mu$m/min. By the time macroscopic coarsening was first observed, about 7 hours after starvation ($t=0$), \emph{M. xanthus} cells had sped up to $\sim2.5\mu$m/min (Fig.~\ref{fig:TrackingActive}A). We observed that wild-type cells became more active over the time course of our experiment (Fig.~\ref{fig:activecells_vdist}), in agreement with previous reports of an initial ``resting'' phase upon starvation~\cite{jelsbak2002pattern}
\footnote{$\Delta$FrzE cells do not change speed when starving, indicating a link between the Frz pathway and gliding speed.}
. We also found that reversal frequency decreased from 8.5~h$^{-1}$ to 5~h$^{-1}$ over the coarse of the  experiment, with most of the reduction occurring  before $t=0$ (Fig.~\ref{fig:TrackingActive}B).

A combination of increased $v_0$ and decreased $f_\text{rev}$ produces a reduction in \Pe$^{-1}_\text{r}$ from $\sim1.8$ before starvation to $\sim 0.25$ when mature fruiting bodies have been formed (Fig.~\ref{fig:TrackingActive}C). Most of this reduction occurs in the first 7 hours after starvation, and  \Pe$^{-1}_\text{r}$ plateaus at the same time as the onset of macroscopic coarsening ($t=0$, Fig.~\ref{fig:PhaseDiagram}). The apparent critical \Pe$^{-1}_\text{r} \sim 0.3$ lies in the region of phase space that we estimate to include the spinodal line. We note that the phase diagram in Figure~\ref{fig:PhaseDiagram} is made by examining the behavior of non-reversing cells that do not alter their behavior over time (Fig.~\ref{fig:TrackingActive}A). The correspondence between the critical \Pe$^{-1}_\text{r}$ at which starving wild-type  cells start to undergo macroscopic coarsening and the estimated spinodal line derived from $\Delta$FrzE cells strongly indicates that \Pe$^{-1}_\text{r}$ is sufficient for understanding the initial process of phase separation.

To further test this idea, we added nigericin to starving WT cells and found that above a concentration of 2.5$\mu$M, the population was unable to form FBs. This is likely due to the inability of these cells to speed up enough to cross the critical \Pe$^{-1}_\text{r}$. Finally, it is possible that {\em M. xanthus} cells could alter their reversal frequency directly as a function of cell density, explaining the observed reduction in $f_\text{rev}$ over time. However, we measured reversal frequency for isolated cells and did not observe any significant change in reversal frequency (Fig.~\ref{fig:Rev_hist}).

Here we present a simple physical picture of {\it M. xanthus} FB formation based on the statistical physics of active populations. Before starvation, cells move slowly and reverse frequently, favoring a homogeneous population on the surface. Upon starvation, wild-type cells speed up and reverse less often, producing a situation favorable for phase separation and FB formation. However, fruiting body formation is ultimately a sophisticated biological process and it is likely that our simple model does not capture the entirety of its complexity. Many of the details that we purposely left out of our analysis could play a role in the specific evolution and shape of the final fruiting bodies. These include cell-cell alignment, the effects of ``slime following,'' and cell-cell communication via the C- and A-signaling mechanisms. For example, $\Delta$FrzE cells do not form stable fruiting bodies. While initially stable, droplets typically fall apart at the end of 24 hours, towards the end of traditional FB development. This potentially indicates that additional biological or chemical mechanisms could play a role in FB stability over long times. More complicated models of {\it M. xanthus} aggregation may uncover the role of these additional parameters~\cite[see e.g. ref][]{cotter2017data}, but it is unlikely that they will change the basic features we have observed here. 

The two-dimensional ABP model we used differs from FB formation in several important ways. ABPs phase separate via a jamming aggregation process where the particles slow down due to crowding, whereas {\it M. xanthus} cells remain motile throughout the process of FB formation (see e.g. Movie S3 and S6). Importantly, the fruiting bodies are three-dimensional structures that appear to be 'dewetted' from the initial homogenously spread quasi-two-dimensional layer of cells. However, the similarity in the scaling exponents for coarsening and the phase diagrams may potentially indicate that these processes, while seemingly very different on the microscopic scale, may in fact belong to the same universality class of active systems. Future work tracking cells and monitoring droplet shape in three dimensions should lead to a more accurate theory of the phase separation which might be viewed as the dewetting of an active bacterial fluid layer into 3D droplets.

The authors thank Suraj Shankar and Lisa Manning for useful discussions and Suraj Shankar for the calculation of the MSD for ABPs with reversals. MCM was supported by NSF-DMR-1609208 and Simons Foundation Targeted Grant in the Mathematical Modeling of Living Systems 342354. MCM and AP acknowledge support by the NSF IGERT program through award NSF-DGE-1068780. MCM, AP and DY were additionally supported by the Soft Matter Program at Syracuse University. Simulations were carried out on the Syracuse University HTC Campus Grid supported by NSF-ACI-1341006. DY acknowledges partial support by from MINECO (Spain) and FEDER (European Union) FIS2015-65078-C2-1-P. JWS, ST, and GL were supported by NSF-PHY-1401506, NSF-PHY-1521553, the Center for the Physics of Biological Function NSF-PHY-1734030, and an HFSP Cross Disciplinary Fellowship to ST. RDW and FB were supported by NSF-DBI-1244295. Part of this work was performed at the Aspen Center for Physics, which is supported by NSF-PHY-1607611.

\bibliographystyle{apsrev4-1}
\bibliography{Liu_PRL2018}

\clearpage
\newpage
\onecolumngrid

\beginsupplement
\setcounter{page}{1}

\begin{center}
{\Large {\bf Supplementary Material}}
\end{center}

\section{Methods}
\subsection{Bacterial strains, growth conditions, and developmental experiments}
Liquid cultures of wild-type {\it M. xanthus} strain DK1622 and
$\Delta$FrzE were grown at 32$^\circ$C in agitating CTTYE medium (1.0\% Casitone, 0.5\% yeast extract, 10.0~mM Tris-HCl at pH~8.0, 1.0~mM $KH_{2}PO_{4}$, and 8.0~mM $MgSO_{4}$). Kanamycin (40~$\mu$g/ml) was added to liquid cultures of $\Delta$FrzE. Starvation assays were performed using non-nutritive Tris phosphate medium (TPM) agarose (10.0~mM Tris-HCl at pH~7.6, 1.0~mM $KH_{2}PO_{4}$, 8.0~mM $MgSO_{4}$, and 1.5\% agarose). To induce fruiting body (FB) development, growing cells were harvested from liquid culture at mid-log phase and resuspended to a final concentration of various densities in TPM: $5\times10^{7}, 1.5\times10^{8}, 2.5\times10^{8}, 5\times10^{8}, 2.5\times10^{9}$~cells/mL. 10 $\mu$l spots were plated on a TPM agarose slide complex and allowed to dry as described previously~\cite{bahar2014describing,thutupalli2015directional}. To modulate velocity, cells suspensions and TPM agarose was supplemented with nigericin sodium salt at concentrations of 0, 1, 2, 4, and 10~$\mu$M.

\subsection{Imaging and tracking}
\begin{figure*}[h!]
\includegraphics[width=.35\textwidth]{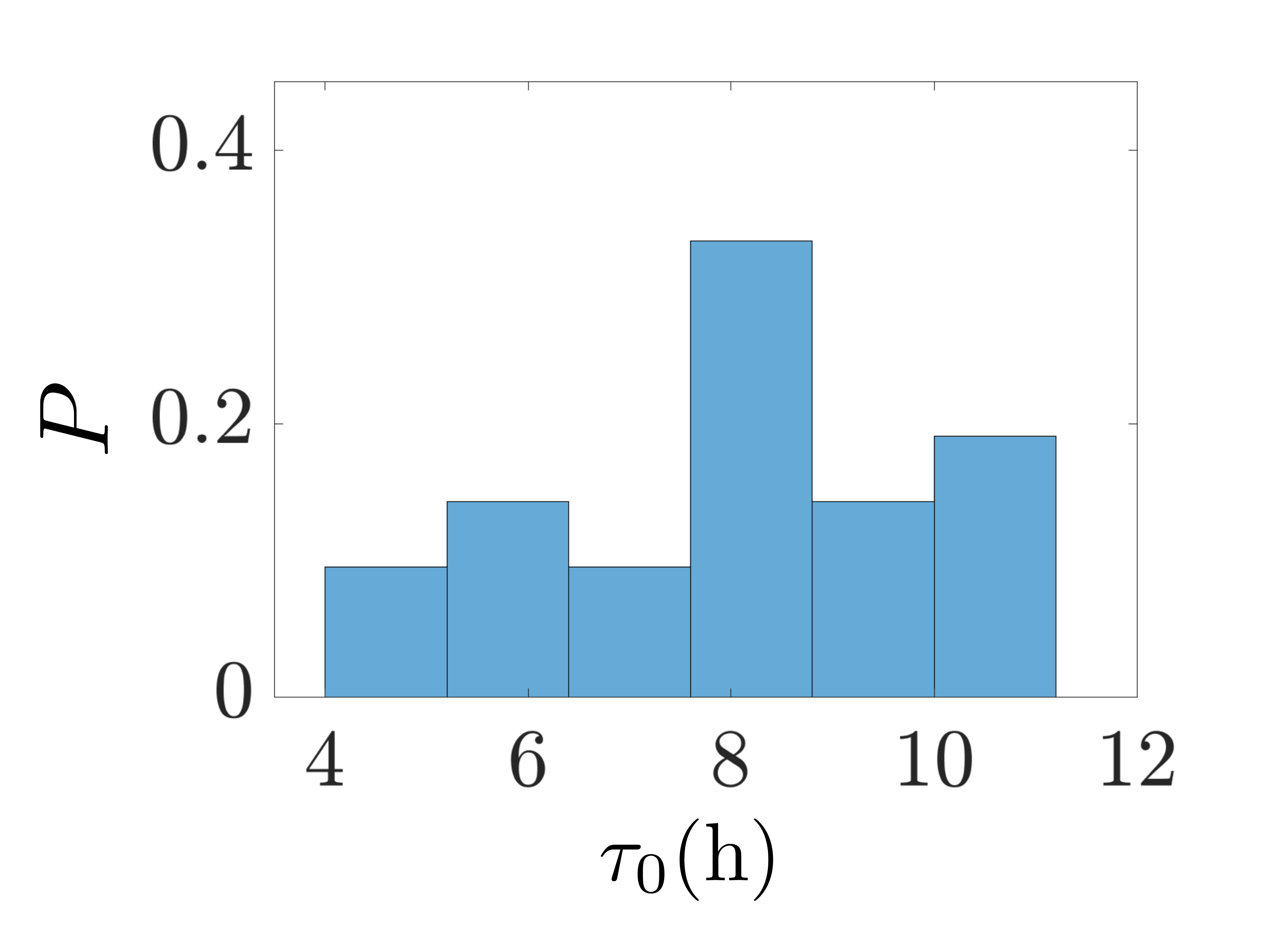}
\caption{Probability density of times between inoculation and the first evidence of macroscopic coarsening (defined as $t=0$ in the text).
}
\label{fig:tauHist}
\end{figure*}

Cells were imaged at 100$\times$ and 20$\times$ magnification to record the behavior of both single cells and aggregates, respectively. For 100$\times$ magnification experiments, cells were imaged on a modified Nikon TE2000 inverted microscope with an oil-immersion objective (NA~1.49). Details of this imaging setup and auto-focusing strategy were reported previously~\cite{thutupalli2015directional,sabass2017force}. Images were recorded at a rate of one frame per 10 seconds with single cell resolution. For 20$\times$ magnification experiments, bright-field movies were taken with phase contrast on a Nikon microscope with 1 minute time resolution. To measure the length-scale evolution during phase separation, we used a 20$\times$ magnification home-built bright field microscope at a rate of one frame every 10 seconds for 24 hours. For wild-type starvation experiments, we visually inspected each movie to identify the time, $\tau_0$, at which macroscopic coarsening began. Across experiments, $\tau_0$ ranged from 4 to 11 hours, with an average of $8.1 \pm 1.9$~hours (Fig.~\ref{fig:tauHist}). For each movie, this time was labeled as $t=0$, with negative times occurring before the onset of coarsening and positive times occurring after.

\begin{figure*}[h!]
\includegraphics[width=.7\textwidth]{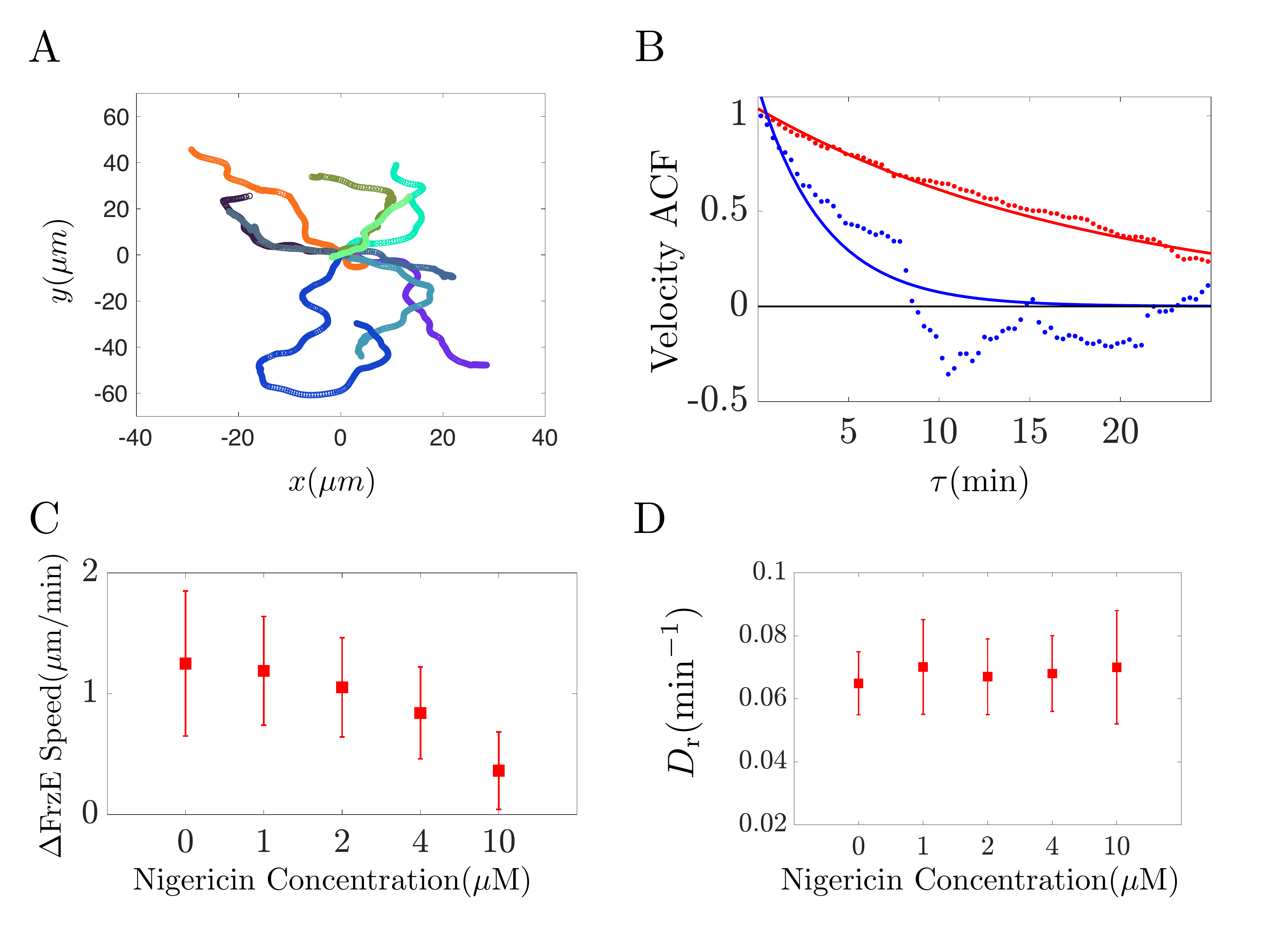}
\caption{Isolated cell tracking. (A) Ten example tracking of isolated $\Delta$FrzE cells. Each track starts at position $(0,0)$.  (B) Average velocity autocorrelation function (ACF) for WT  (blue, N=29) and $\Delta FrzE$ (red, N=32) in absence of nigericin. The rotational diffusion constant $D_\text{r}^\text{eff}$ is equal to the temporal decay constant calculated from the fit to a single exponential function (solid lines). (C) Speed of $\Delta$FrzE cells in the presence of various nigericin concentrations. $\sim$30 cells are used to calculated each average and standard error. (D) Rotational diffusion coefficient of the non-reversing mutant calculated at each nigericin concentration.
}
\label{fig:experiment_track}
\end{figure*}

Cell tracking using 100$\times$ bright-field images was performed using the previously published BCTracker algorithm~\cite{thutupalli2015directional}. Cell velocities were calculated as the spatial displacement of the tracked centroid of each cell per frame divided by the frame rate. Because of noise in the tracking, we slightly overestimate the speed of very slow cells but this small effect does not affect the conclusions in the paper. Additionally, the speed shown for $\Delta$FrzE shown in Fig.\ref{fig:experiment_track}C is corrected by taking into account the fact that only 75\% of cells in our field of view were motile.

The rotational diffusion coefficient ($D_\text{r}^\text{eff}$) was calculated from the decay time of the velocity autocorrelation function from isolated cells. The autocorrelation functions from $\sim$30 cells were first averaged together and then fit to a single decaying exponential (Fig.~\ref{fig:experiment_track}). Only tracks with a trajectory longer than 25$\mu$m were included in this analysis. 

The mean square displacement (MSD) versus time for both wild-type and the non-reversing cells was calculated using low density, high magnification tracking data. Time was rescaled by the $D_\text{r}^\text{eff}$ calculated from the velocity autocorrelation functions. Upper and lower bond of the shaded uncertainty area were calculated as one standard error of the mean when considering uncertainty in $v$ and $D_\text{r}^\text{eff}$.

\begin{figure}[h!]
\includegraphics[width = .7\columnwidth]{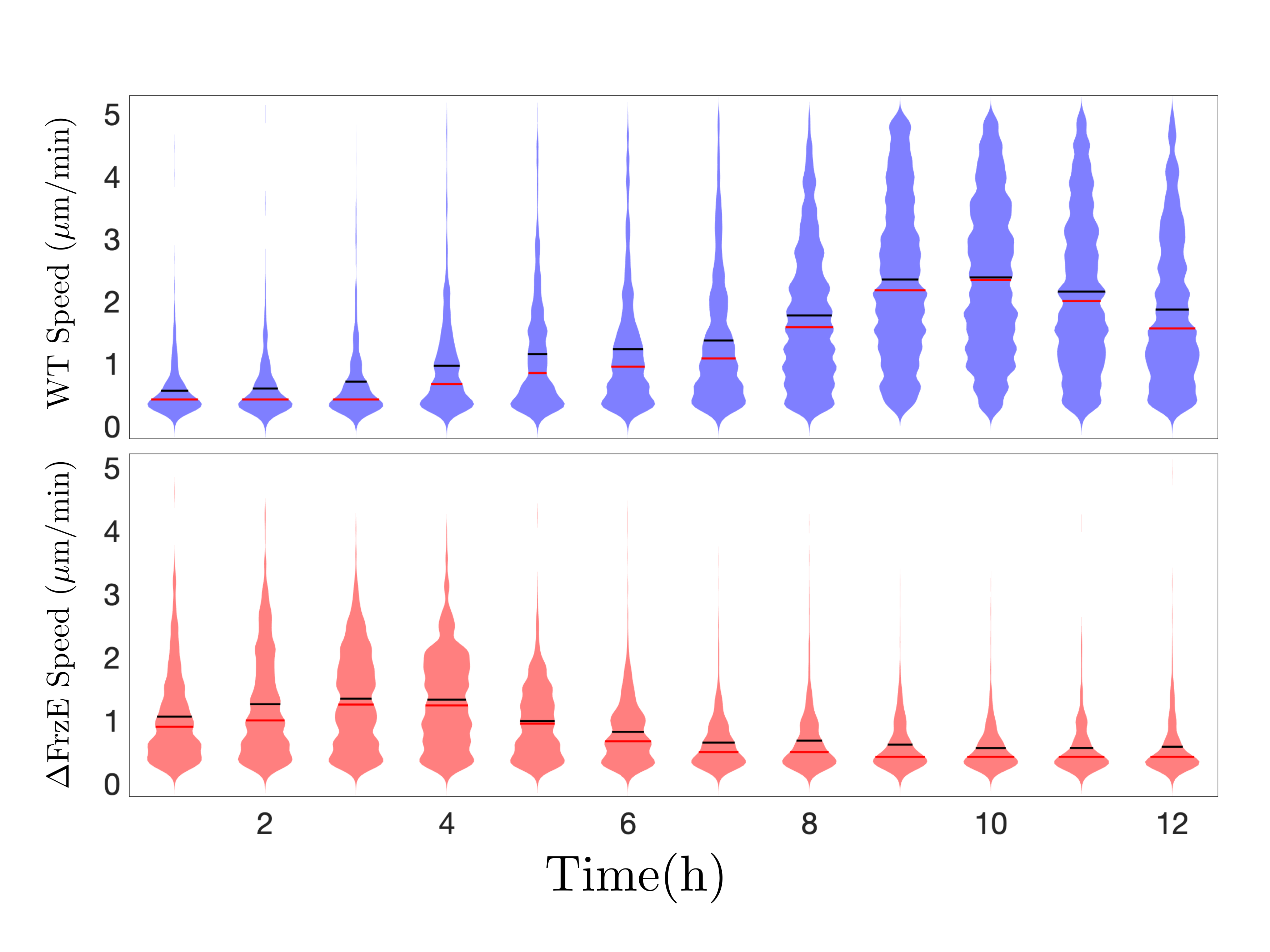}
\caption{Speed distributions for wild-type (blue) and the non-reversing mutant (red). Mean values are indicated with a black line; median values with a red line.}
\label{fig:activecells_vdist}
\end{figure}

To study cell motility in dense populations where phase separation occurred, we mixed fluorescently-labeled cells with unlabeled cells at ratio of 1:400 and tracked the fluorescent signal over time. Both wild-type and $\Delta$FrzE cells were labeled with Alexa Fluor 594 carboxylic acid succinimidyl ester. To stain cells, cells were grown to mid-log phase, harvested by centrifugation and resuspended in MC7 buffer. 2~$\mu$l of dye (10 mg/ml, dissolved in DMSO) and 5$\mu$L of 1M $NaHCO_{2}$ were added to 100 $\mu$l of cells and shaken vigorously at 100 RPM for 1 hour in the dark at room temperature. Cells were then pelleted by centrifugation, washed 3 times in TPM and microscopically examined. Fluorescent microscopy images were taken at a rate of one frame per minute for the first 15 min in each hour to minimize the amount of laser exposure for cells. Experiments lasted 11 hours in total. Fluorescent cells were tracked using a particle tracking algorithm developed by Crocker, Grier and Weeks~\cite{crocker1996methods}, and cell velocity was calculated in the same manner as for the high magnification movies. Detailed speed distributions for WT and $\Delta$FrzE are shown in Figure~\ref{fig:activecells_vdist}.

\begin{figure}[h!]
\includegraphics[width = .7\columnwidth]{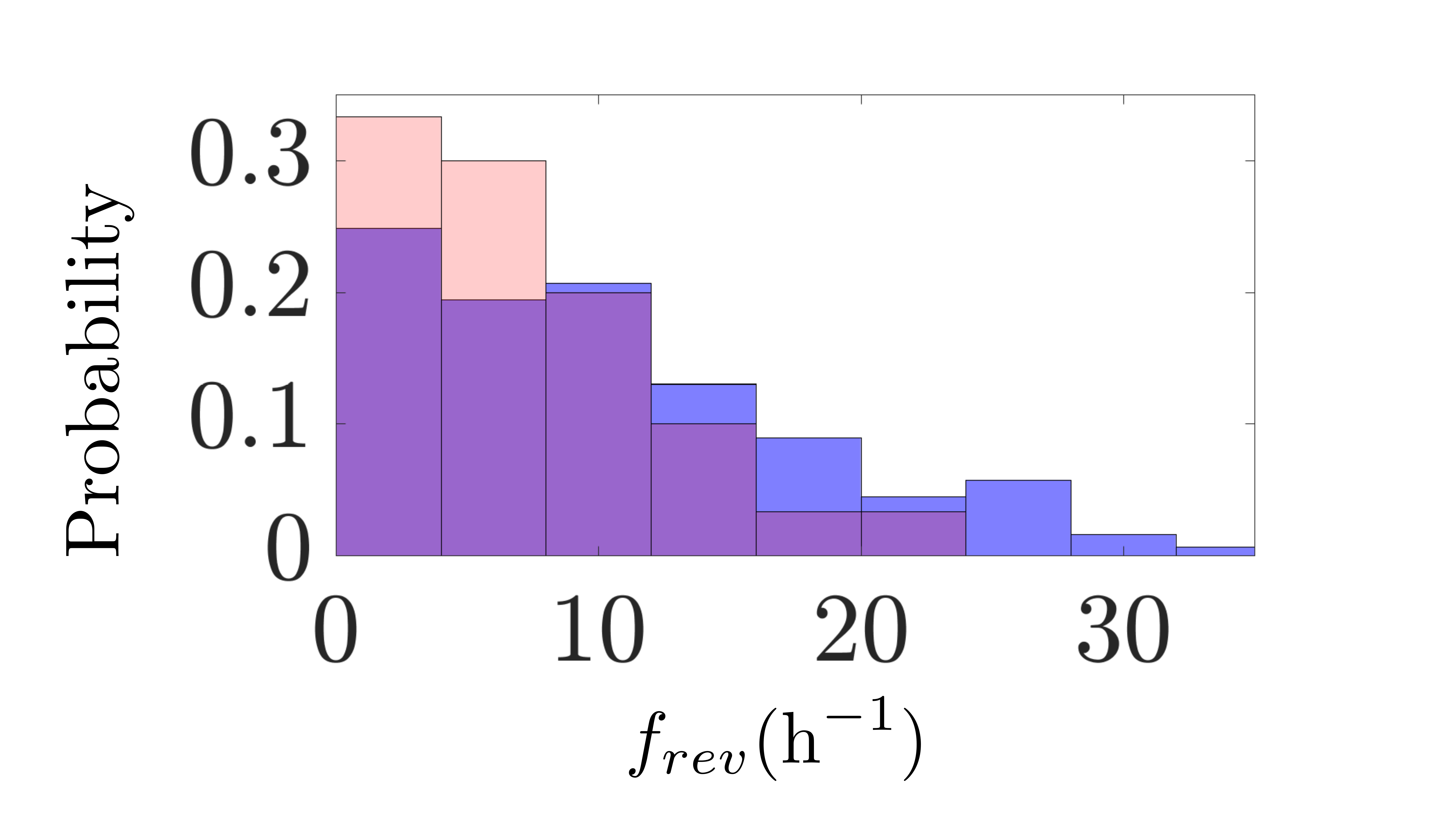}
\caption{Distribution of reversal frequencies for wild-type cells in groups at high density (blue) and in isolation at low density (red). Wild-type {\it M. xanthus} cells do not show significantly different reversal frequency in isolation versus in groups. The mean reversal frequency for isolated cells is $6.28 \pm 1.49$ h$^{-1}$ and for cells in groups is $8.58 \pm 2.36$ h$^{-1}$.  Kolmogorov-Smirnov test yields $p = 0.10$}.
\label{fig:Rev_hist}
\end{figure}

We calculate directional reversals by examining changes in cell velocity. A reversal event in both tracking methods was recorded when the two-dimensional dot product of the velocity vectors between two successive time points was negative. We counted the number of reversal events for each bacterium during the length of its track to calculate its reversal frequency, and then averaged the reversal frequency for all tracked cells. This analysis was done separately for each hour for the duration of an experiment to estimate the change in $f_\text{rev}$ over time. We do not include the reversal frequencies from the first hour of the experiments shown in Fig.~\ref{fig:TrackingActive}B because cell speed is very low during this time and we overestimate the reversal frequencies due to tracking noise.

At high cell densities, it is experimentally difficult to break up cell clumps that have formed in the liquid culture. Thus, at the beginning of a movie we sometimes see isolated clumps of cells that are not fruiting bodies and which dissolve within the first hour of experiment as cells migrate out of them. The locations of these cell clumps are uncorrelated with the eventual positions of the aggregates and fruiting bodies, and we conclude that clumps are unrelated to fruiting body formation.

\subsection{Simulation details and parameters}

\begin{figure}[h!]
\centering
\includegraphics[width=.7\columnwidth]{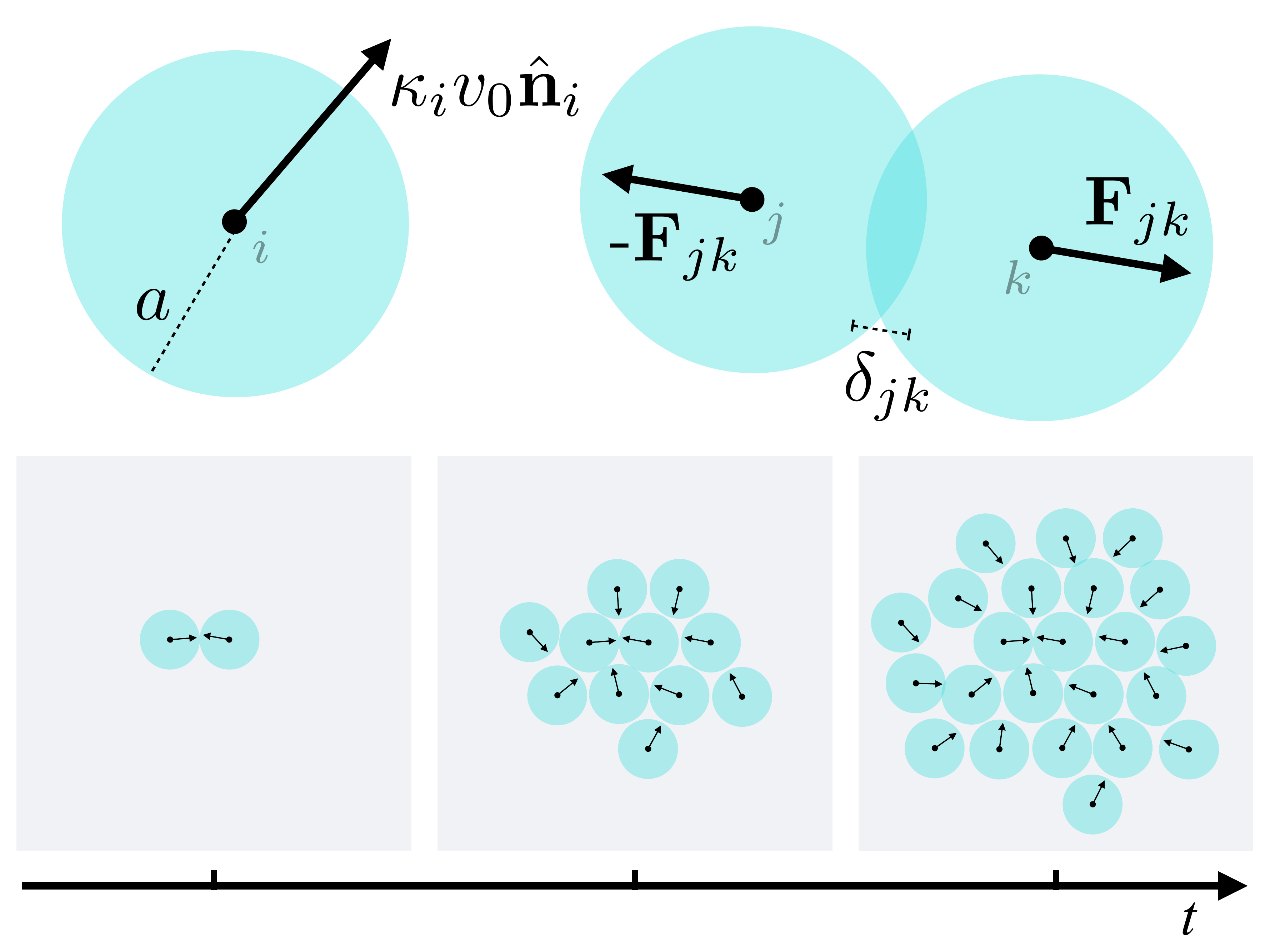}
\caption{Schematic of a minimal reversing ABP model. 
\label{fig:simcartoon}}
\end{figure}

\begin{figure}[h!]
\includegraphics[width = .7\columnwidth]{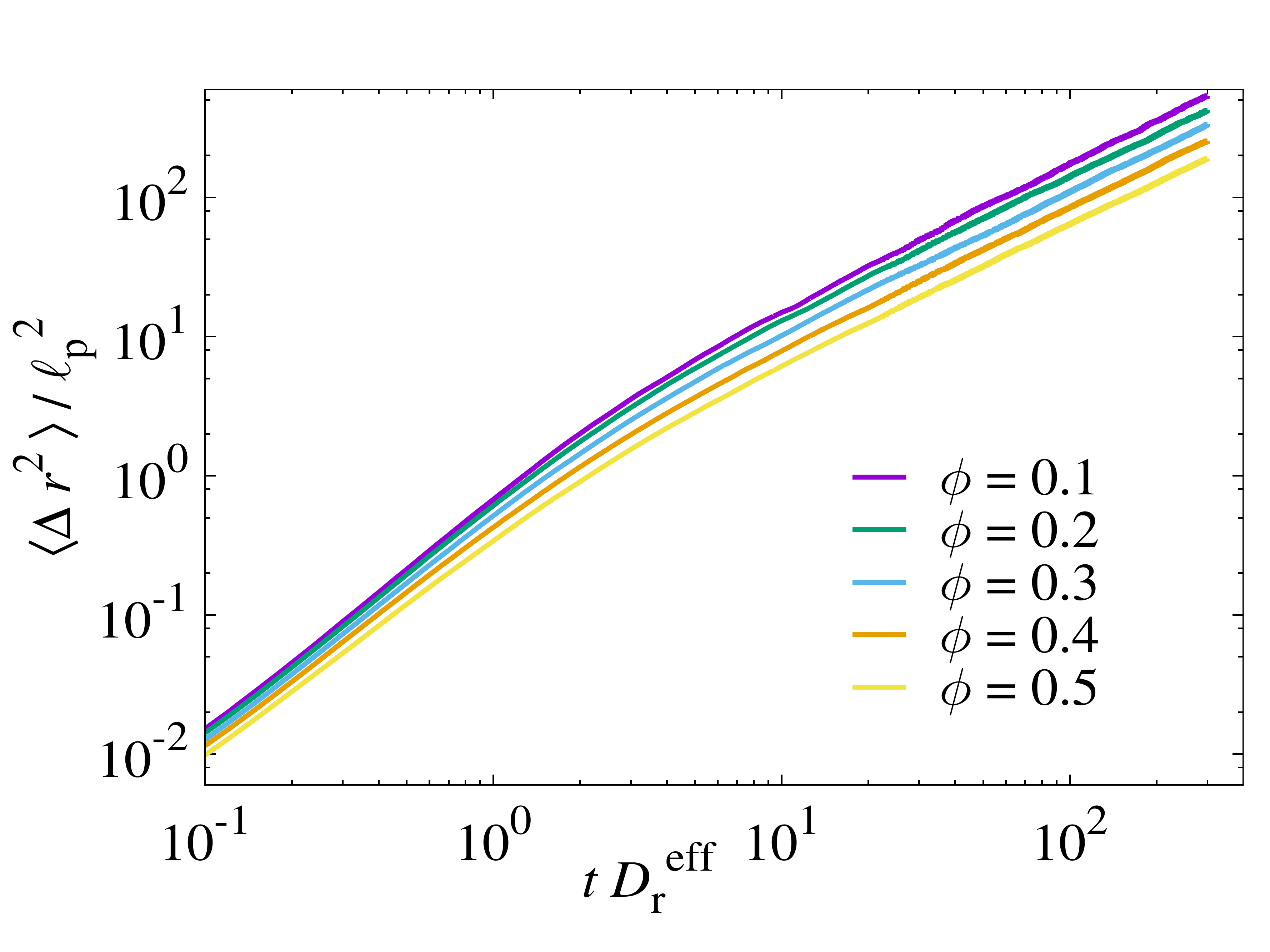}
\caption{MSD for finite simulation densities. Collisions reduce the particles' displacement, but the crossover from ballistic to diffusive motion is still controlled by $\tau_\text{r}^{-1}$. }
\label{fig:MSD_density_sim}
\end{figure}

Each reversing ABP is modeled as a disk of radius $a$, with dynamics governed by over-damped Langevin equations of motion (Fig.~\ref {fig:simcartoon}),
\begin{align}
\dot{\boldsymbol{r}}_i &= v_0 \kappa_i (t) \hat{\boldsymbol{n}}_i + \mu \sum_j \boldsymbol{F}_{ij}\;, &
\dot{\theta}_i &=\sqrt{2D_\text{r}} ~\eta_i(t) \;, \label{eq:motion_r}
\end{align}
\noindent where $\boldsymbol{r}_i$ and $\hat{\boldsymbol{n}}_i =\left(\cos\theta_i,\sin\theta_i\right)$ are the position and orientation of the $i^\text{th}$ disk. \eqref{eq:motion_r} describes the velocity of the $i^{\text{th}}$ particle as a function of its self-propulsion and steric interactions. The direction of self-propulsion is updated stochastically according to a random torque $\eta_i(t)$ of unit variance. In~\eqref{eq:motion_r} We modify the standard ABP model by incorporating directional reversals through a function $\kappa_i(t)$, which takes the values $\pm1$, changing sign 
at times given by a Poisson process with a mean reversal frequency $f_\text{rev}$. The force $\boldsymbol F_{ij}$ is purely repulsive and represents an excluded-volume interaction. We use a harmonic potential, with $\boldsymbol{F}_{ij} = k ( 2 a - r_{ij} ) \hat{\boldsymbol{r}}_{ij}$ for $r < 2a$ and $F_{ij} = 0$ otherwise, where $\hat{\boldsymbol{r}}_{ij}=\boldsymbol{r}_i-\boldsymbol{r}_j/|\boldsymbol{r}_i-\boldsymbol{r}_j|$

We simulated Equation~\eqref{eq:motion_r} using a standard Brownian Dynamics algorithm in an $L\times L$ box with periodic boundary conditions. In all cases, we use $k = \mu = 1$ so the interaction timescale $\tau_\text{D} =(\mu k)^{-1}$ sets the unit time and we use the particle radius $a$ as the unit of length ($a = 1$). To prevent particles from passing through each other, we set $v_0 = (a \mu k)/100$. We fix the packing fraction $\phi = N \pi a^2/L^2$, which sets the total number of particles, $N$. The rotational diffusion $D_\text{r}$ and the reversal frequency $f_\text{rev}$ are varied to obtain the desired $\Pe_\text{r}^{-1}$.

The integration timestep $\Delta t$ has to be kept very small with respect to
$\tau_\text{r}=1/D_\text{r}^\text{eff}$, which, together with our choice of a
smooth spring potential for the repulsion, means that the particle positions
can be safely updated with a simple Euler method.  There are two choices for
the reversals. One option is to consider them in an event-driven way, i.e., generate
a reversal time $t_\text{rev}$ from a Poisson distribution with
$\lambda=1/f_\text{rev}$ and then integrate until
time $t_\text{rev}$ is reached, at which point the orientation of the velocity
is reversed and a new $t_\text{rev}$ is generated. However, since $\Delta
t\ll 1/f_\text{rev}$ we have instead checked for reversals at each time step
(i.e., changing the sign of the velocity with probability $\Delta t f_\text{rev}$).
This method is, in principle, marginally less efficient, since it requires more
random numbers, but this operation is computationally negligible compared with
the evaluation of particle forces. Finally, in order to represent the
rotational diffusion one only has to evaluate $\theta_i(t+\Delta t) =
\theta_i(t) + \zeta_i(t)$, where $\zeta_i$ is a Gaussian random variable with
zero mean and $\sigma = \sqrt{2D_\text{r}\Delta t}$.

For each set of parameters, we average over $10$--$100$ runs and use a jackknife method~\cite{Amit2005,Patch2017} to estimate statistical errors. In order to compute the length scale $L(t)$ and its coarsening exponent (Fig.~\ref{fig:SnapShots}D) we used a large system size with $L=1000$, $\Pe_\text{r}^{-1}=0.01$ and a packing fraction of $\phi=0.5$ ($N=159, 154$ particles), averaging over $100$ independent runs. For other quantities we did not need such a large system size. The phase diagram was computed on systems with $L=200$ ($10$ runs for each set of parameters).

\subsection{\boldmath Measuring $L(t)$}

\begin{figure}[h!]
\begin{center}
\includegraphics[width=0.7\columnwidth]{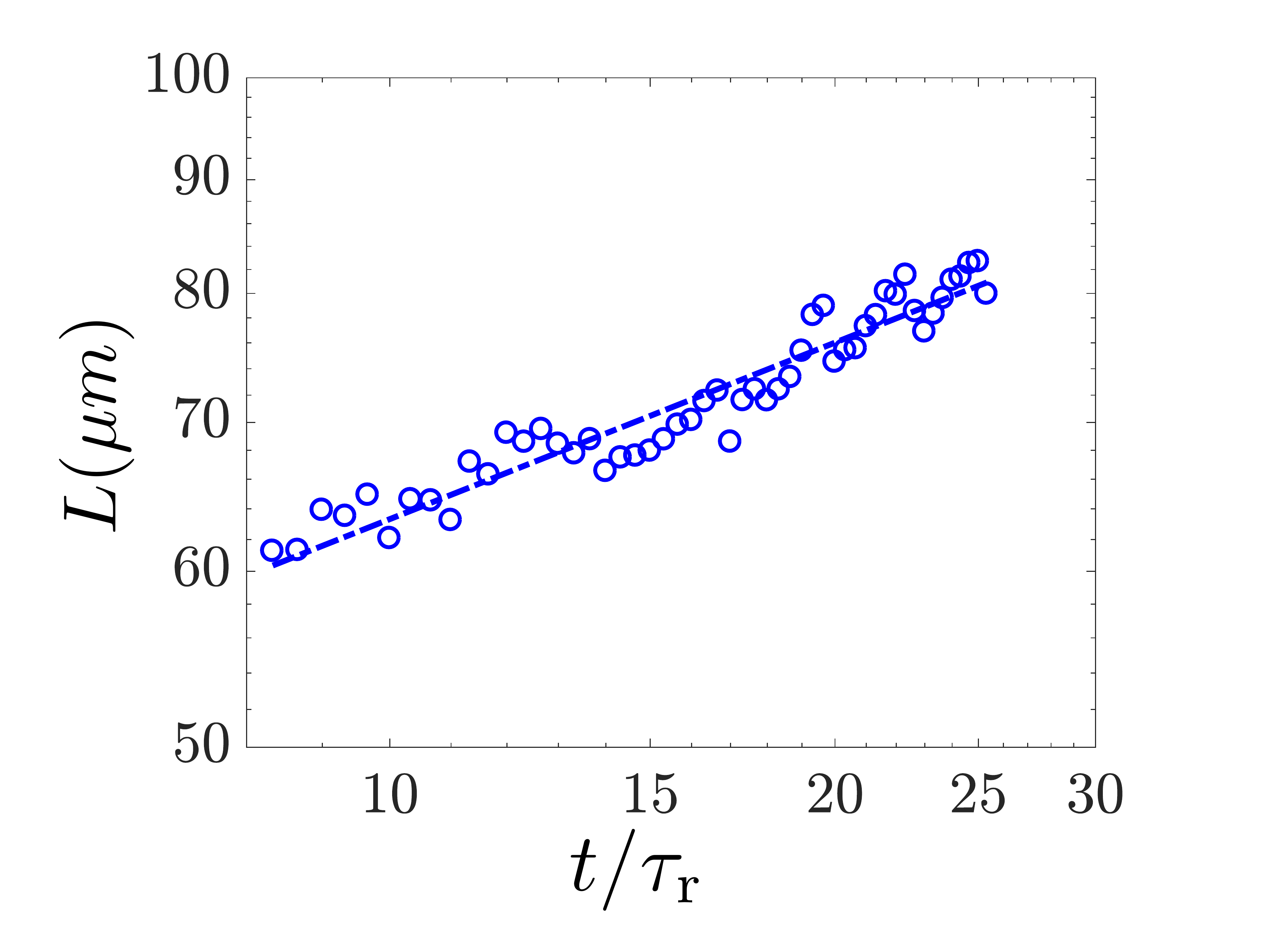}
\end{center}
\caption{ Coarsening behavior of non-reversing mutant. Similar scaling to wild-type (Fig.\ref{fig:SnapShots}D) is found, with an exponent of $\alpha_{\Delta\text{FrzE}} = 0.27 \pm 0.02$. Time is rescaled as described with $\tau_{\text{r}} = 10$ min. }
\label{fig:frzE_evolution}
\end{figure}

We estimated local density in our movies as the intensity of measured light intensity at each pixel, where density has an inverse relationship with light intensity (darker is more dense). For cases of nucleation and growth in wild-type {\it M. xanthus}, $L(t)$ was measured as the averaged over 7 individual fruiting bodies. Time $t = 0$ for each fruiting body was estimated when the nuclei was first visible. For cases of spinodal decomposition, we calculated the radial component of the Fourier transform of each frame as the experimental $S(k)$ distribution. The radial Fourier transform was then fit to the sum of a Gaussian function and an exponential decay to account for the inhomogeneous lighting effects in the experiments. In our analysis, we fitted conditions where no visible {\it M. xanthus} structures were visible and found that the decay constant in the exponential function has a length scale $= 358.3 \mu$m. This constant was then fixed for all other experiments and the mean wave number extracted from the Gaussian function was taken to be the inverse of the dominant length scale in each frame. For shorter times ($<100$ min), it is difficult to pick out a distinctive peak position from the Fourier transform. Thus we only present scaling form $t/\tau_{\text{r}} > 10$. Non-reversing cell populations also coarsen upon starvation with the same temporal scaling as the wild-type, although this only occurs for a short time as these droplets are unstable (Fig.~\ref{fig:frzE_evolution}).

\begin{figure}[h!]
\begin{center}
\includegraphics[width=0.7\columnwidth]{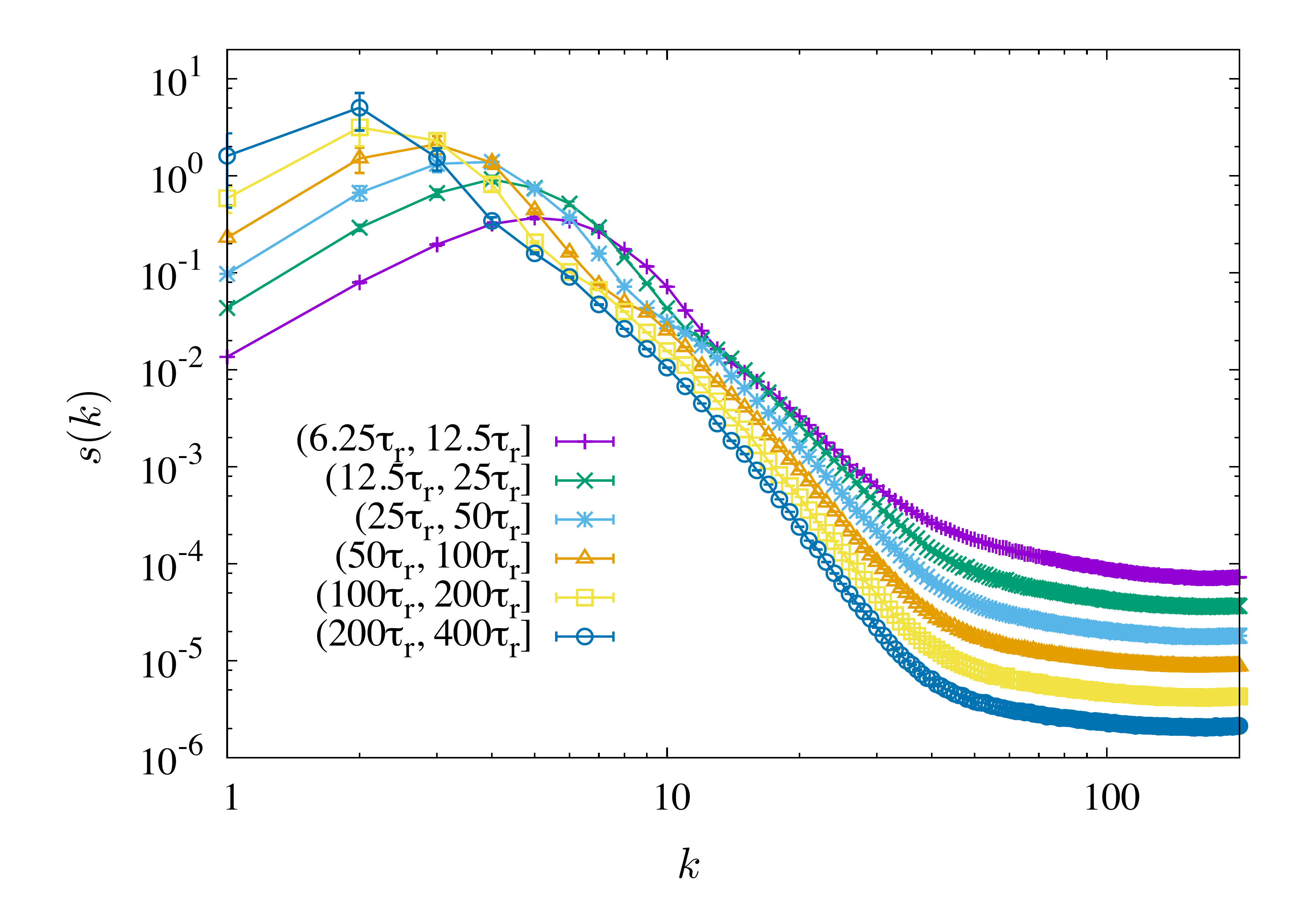}
\end{center}
\caption{Static structure factor calculated for time intervals used to produce $L(t)$ ($\phi=0.5$,$Pe_\text{r}^{-1} = 0.01$, $L=10^3$). This is an example of spinodal phase decomposition, where this distribution is expected to tighten around smaller $k$ as dense regions coarsen with time.}
\label{fig:sk_evolution}
\end{figure}

In simulations, we discretize the system by dividing it into bins with width equal to the particle radius $a$ and assigne a binary 1 or 0 value to each bin depending on whether or not a particle is centered in it. In order to reduce noise, the resulting density distribution is averaged over exponentially increasing temporal bins \cite{Patch2017}. The FFT is then computed in 2D to produce the structure factor $S(\boldsymbol k,t)$ (Fig.~\ref{fig:sk_evolution}). The average length scale was calculated from the first moment of $S(k,t)$ 
\begin{equation}
L(t) = \frac{\int_{2\pi/N_\text{FFT}}^{k_\text{max}} S(k,t)\ \mathrm{d}k}{\int_{2\pi/N_\text{FFT}}^{k_\text{max}} k S(k,t)\ \mathrm{d}k}\;,
\end{equation}
\noindent where $k_\text{max}$ is chosen to exclude noisy, high-frequency modes and sample the mean of the lowest-wavenumber peak in $S(k)$. We use a jackknife method~\cite{Amit2005,Patch2017} to estimate errors.

\subsection{Distinguishing phase behavior}

\begin{figure*}[h!]
\begin{center}
\includegraphics[width=1\textwidth]{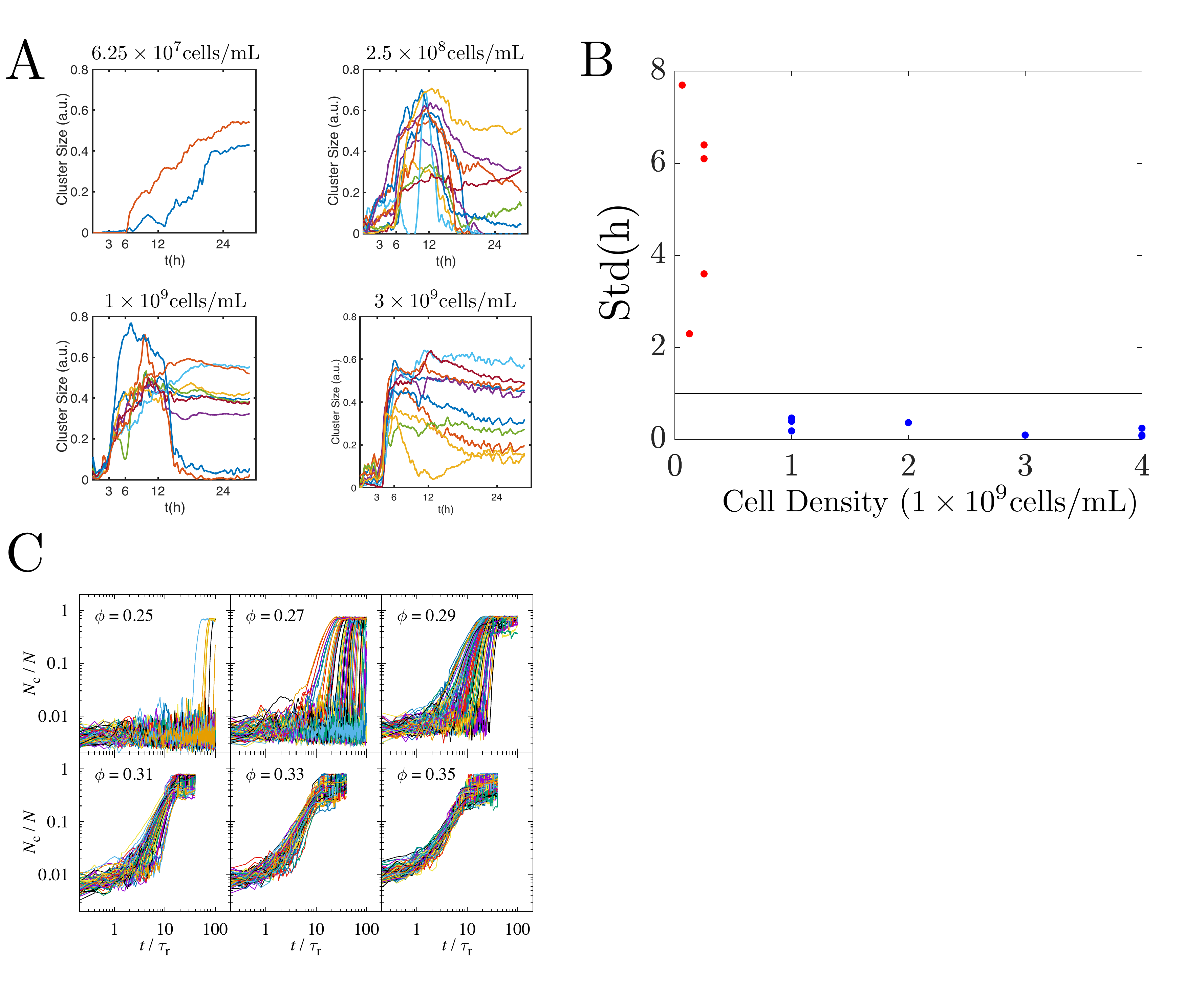}
\end{center}
\caption{(A) Examples of WT FB size evolution at different experimental densities. The top two conditions display nucleation and growth as fruiting bodies form at varying times. The bottom two conditions show spinodal decomposition where the emergence of fruiting bodies is synchronized. (B)  Standard deviation in the time when fruit bodies grow to half of their maximal size versus inoculation density. Red dots represent the system undergoing nucleation and growth and blue dots represent spinodal decomposition. We use a threshold of $\sigma=1\text{ hour}$ to distinguish between the different kinetic mechanisms (black line). (C) The simulated fraction of cells in clusters observed over time is shown for 100 simulations at varying packing fractions ($Pe_\text{r}^{-1}=0.01$). The resulting deviation of the nucleation times shows how as density is decreasing from just inside the coexistence region, significantly longer mean times and wider variances in times of nucleation occur. This provides a clear distinction between spinodal decomposition and nucleation and growth, allowing for a
quantitative threshold between the two.} \label{fig:NGvsSD} 
\end{figure*}

To quantitatively distinguish between nucleation and growth phenomena and spinodal decomposition, we take advantage of the temporally spontaneous nature of the spinodal decomposition phase transition. In spinodal decomposition, as microscopic fluctuations are  unstable, FBs form everywhere throughout the field of view at the same time. In nucleation and growth, the time at which a microscopic density fluctuation becomes large enough to seed a FB is broadly distributed so that FBs form at random times throughout the field. We measured the size of aggregates over time (Fig.~\ref{fig:NGvsSD}A) and then calculated the distribution of times at which aggregates reach one half of their final sizes.
The standard deviation in the time when FBs reach half-maximal size is then compared across experiments at different cell densities (Fig.~\ref{fig:NGvsSD}B). We then set a threshold at $\sigma=1\text{ hour}$ to distinguish different phase transition behaviors.

We performed a similar calculation using the simulations. Since we are limited by the simulation size, in many cases we only get one cluster at the end. To probe if a system undergoes a spontaneous transition for each set of parameters, we performed 100 simulations and compared the cluster growth across simulations (Fig.~\ref{fig:NGvsSD}C). For sets of parameters that lead to spinodal decomposition, the distribution of aggregation times has a clear peak, while for nucleation and growth a large tail develops (with an eventually diverging average as we approach the phase boundary).

\subsection{Phase diagrams}

The experimental phase diagram was produced with the non-reversing mutant $\Delta$FrzE. Experimental conditions used and parameters calculated are specified in Table~\ref{tab: PhaseTable}. All motility parameters were measured at very low densities ($<5\times10^{7}$ cell/mL).

\begin{table*}[h!]
\caption{Experimental conditions and parameters used to produce the phase diagram. (SD Density= densities where spinodal decomposion occured.)} \label{tab: PhaseTable}
\begin{tabular}{|c|c|c|c|c|c|}
\hline 
Nigericin($\mu \text{M}$) & Speed($\mu \text{m/min}$)& $D_{r}(\text{min}^{-1}$)& No Aggregation Density($\times 10^{8}$ cell/mL)& SD Density($\times 10^{8}$ cell/mL)& $\text{Pe}^{-1}_{r}$\\
\hline 
0& 1.25& 0.065& 0.5, 1 & 2.5, 4, 5, 25& 0.130\\
\hline

1& 1.19& 0.070& 0.5, 1 & 2.5, 5, 25& 0.147\\
\hline

2& 1.05& 0.067& 0.5, 1 & 2.5, 25& 0.159\\
\hline

4& 0.84 & 0.068& 0.5, 1 & 2.5, 25& 0.202\\
\hline

10& 0.36& 0.070& 4, 5, 25& None & 0.481\\
\hline
\end{tabular}
\end{table*}

\begin{figure}[h!]
\begin{center}
\includegraphics[width=0.7\textwidth]{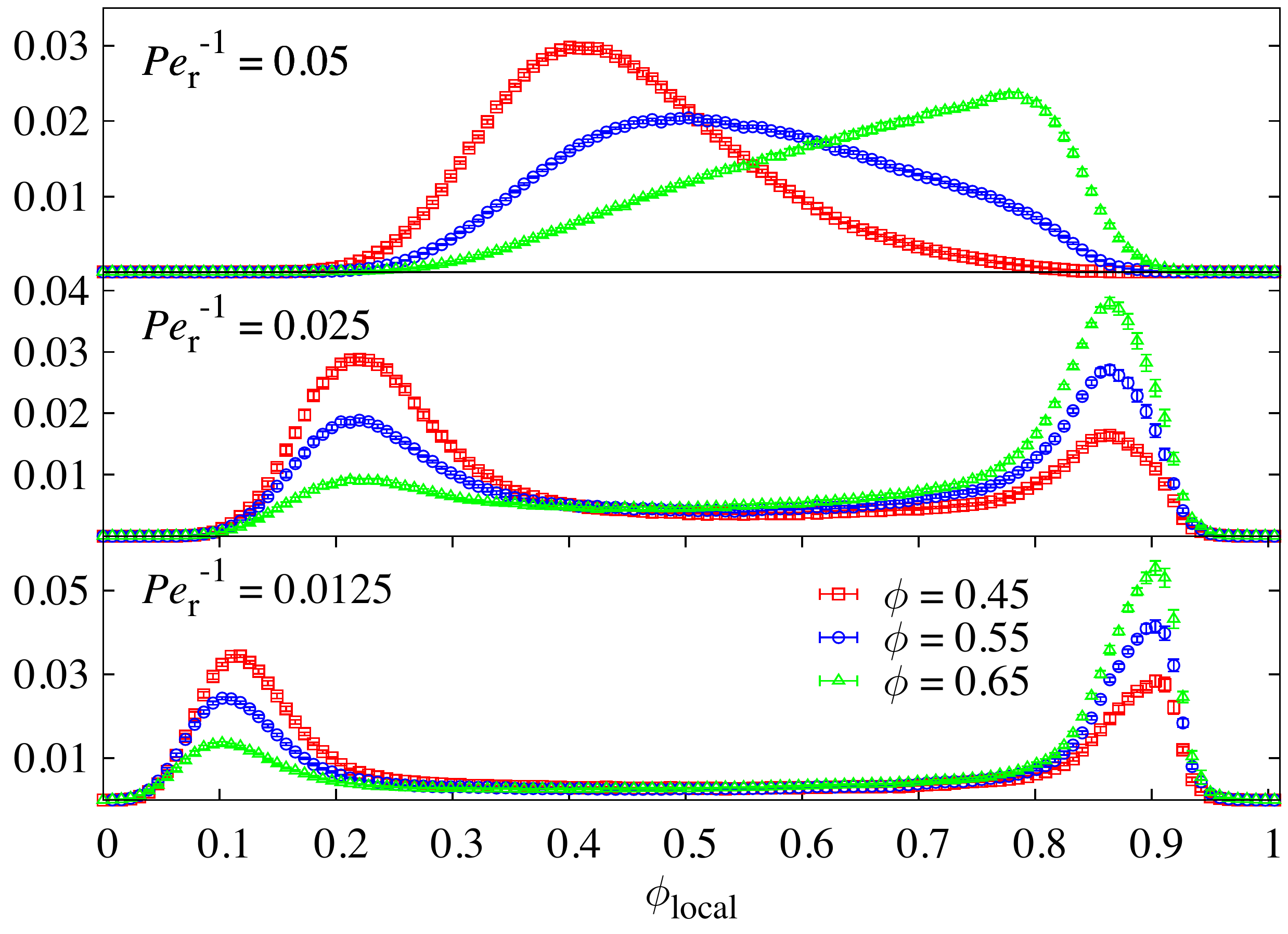}
\caption{
Local density distributions for three different mean densities,
$\phi = \{0.45,0.55,0.65\}$, at three different rotational Peclet numbers,
$Pe_\text{r}^{-1} = \{0.005,0.0025,0.00125\}$. The peaks of these distributions
are used to draw the boundary in the simulation phase diagram.
}
\label{fig:local_density}
\end{center}
\end{figure}

To make the phase diagram from simulations, we follow a simple technique used previously for ABPs by measuring the distribution of local densities in our simulations~ \cite{Redner2013}. In this technique, a unimodal distribution signifies a homogeneous system while a bimodal distribution signifies phase-separated state. Examples
of these are shown in Fig.~\ref{fig:local_density}. We sample local density using square windows of width $L_\text{W} = 20a$ (the total size of the system is $L=200a$). For high density circle-packing, which is at a density above those we generally measure in the dense phase, a window would contain $\approx115$ particles. We use this to set the density bin width of length $d\phi \approx 0.01$.

Drawing from the last half (steady state) of each run, we average histograms
over 10 runs. We determine dominant phase densities by selecting the peaks
from each distribution. These densities are then used as coordinates in our
phase diagram, paired with the configuration $Pe_\text{r}$.

\section{Effective rotational diffusion in the presence of directional reversals}

We consider the over-damped dynamics of a single self-propelled particle with directional reversals in two dimensions. The directional unit vector $\hat{\mathbf{n}} = (\cos\theta, \sin\theta)$ is set by the angular direction $\theta \in [0,2\pi)$. This evolves in time as
\begin{equation}
\text{d}\theta = \sqrt{2 D_\text{r}} \text{d}W_t\; ,
\end{equation}
where $D_\text{r}$ is rotational diffusion and $W_t$ is a Wiener process. Particle direction periodically reverses as a Poisson process where the waiting time distribution between two reversals is exponential. The overdamped equation of motion is expressed as a stochastic process driven by dichotomous Markov noise (DMN)\cite{Sancho1984},
\begin{equation}
\text{d}\mathbf{x}(t) = v_0 \hat{\mathbf{n}}(t) \text{d}\xi_t\; ,
\end{equation}
where $\xi_t \in \{-1,1\}$ is the symmetric DMN with zero mean and exponential correlation $\mathbb{E}[\xi_t,\xi_{t'}] = \exp(-2f_\text{rev}|t-t'|)$, and $f_\text{rev}$ is the mean reversal rate. Particle position $(x(t),y(t))$ evolves as
\begin{align}
x(t) &= x_0 + v_0 \int_0^t \text{d}\xi_{t'} \cos\theta(t') \\
y(t) &= y_0 + v_0 \int_0^t \text{d}\xi_{t'} \sin\theta(t') \;,
\end{align}
where $(x_0,y_0)$ is the initial position. Since $\theta(t)$ and $\xi_t$ are independent random variables, we separately average over all realizations of the rotational noise and initial angular conditions and then average over all realizations of the DMN to calculate the mean-square displacement
\begin{equation}
\langle |\mathbf{x}(t)-\mathbf{x}(t')|^2 \rangle = \frac{v_0^2}{2 D_\text{r}^\text{eff}} \bigg[ t - \frac{1}{D_\text{r}^\text{eff}} \big(1 - e^{-D_\text{r}^\text{eff}t}\big)\bigg]\; ,
\label{eq:revMSD}\end{equation}
where $D_\text{r}^\text{eff} = D_\text{r} + 2 f_\text{rev}$ is the effective rotational diffusion.

\section{Supplement Movies} 
\noindent {\bf Movie S1:} Gas phase of WT {\it M.xanthus} at $2.5\times10^{7}$ cells/mL \\ \\
\noindent {\bf Movie S2:} Nucleation and growth phase of WT {\it M.xanthus} at $1.5\times10^{8}$ cells/mL \\ \\
\noindent {\bf Movie S3:} Spinodal decomposition phase of WT {\it M.xanthus} at $5\times10^{8}$ cells/mL \\ \\ 
\noindent {\bf Movie S4:} Gas phase of ABP simulation at $\phi = 0.1$ \\ \\
\noindent {\bf Movie S5:} Nucleation and growth phase of ABP simulation at $\phi = 0.29$ \\ \\
\noindent {\bf Movie S6:} Spinodal decomposition phase of ABP simulation at $\phi = 0.5$ \\ \\
\noindent {\bf Movie S7:} $\Delta$FrzE cells undergoing spinodal decomposition at $5\times10^{8}$ cells/mL without nigericin \\ \\
\noindent {\bf Movie S8:} $\Delta$FrzE cells don't form aggregates at $5\times10^{8}$ cells/mL with $10\mu M$ nigericin

\end{document}